# Long-distance spin-transport across the Morin phase transition up to room temperature in ultra-low damping single crystals of the antiferromagnet α-Fe$_2$O$_3$

R. Lebrun[1,2,*], A. Ross[2,3], O. Gomonay[2], V. Baltz[4], U. Ebels[4], A.-L. Barra[5], A. Qaiumzadeh[6], A. Brataas[6], J. Sinova[2,7], M. Kläui[2,3,6,*]

1. Unité Mixte de Physique, CNRS, Thales, Université Paris-Saclay, 91767, Palaiseau, France
2. Institut für Physik, Johannes Gutenberg-Universität Mainz, D-55099, Mainz, Germany
3. Graduate School of Excellence Materials Science in Mainz (MAINZ), Staudingerweg 9, D-55128, Mainz, Germany
4. Univ. Grenoble Alpes, CNRS, CEA, Grenoble INP, SPINTEC, F-38000 Grenoble, France
5. Laboratoire National des Champs Magnétiques Intenses, CNRS-UGA-UPS-INSA-EMFL, F-38042 Grenoble, France
6. Center for Quantum Spintronics, Department of Physics, Norwegian University of Science and Technology, Trondheim, Norway.
7. Institute of Physics ASCR, v.v.i., Cukrovarnicka 10, 162 53 Praha 6 Czech Republic

**Antiferromagnetic materials can host spin-waves with polarizations ranging from circular to linear depending on their magnetic anisotropies. Until now, only easy-axis anisotropy antiferromagnets with circularly polarized spin-waves were reported to carry spin-information over long distances of micrometers. In this article, we report long-distance spin-transport in the easy-plane canted antiferromagnetic phase of hematite and at room temperature, where the linearly polarized magnons are not intuitively expected to carry spin. We demonstrate that the spin-transport signal decreases continuously through the easy-axis to easy-plane Morin transition, and persists in the easy-plane phase through current induced pairs of linearly polarized magnons with dephasing lengths in the micrometer range. We explain the long transport distance as a result of the low magnetic damping, which we measure to be $< 10^{-4}$ as in the best ferromagnets. All of this together demonstrates that long-distance transport can be achieved across a range of anisotropies and temperatures, up to room temperature, highlighting the promising potential of this insulating antiferromagnet for magnon-based devices.**

## Introduction

The ultra-fast magnetization dynamics of antiferromagnets (AFMs) are complex due to the multiple sublattices involved, and have so far been studied mostly by neutron scattering experiments[1,2]. The development of THz spectroscopy combined with the burgeoning field of antiferromagnetic spintronics[3,4] has recently generated exciting predictions and first results on the potential exotic dynamics of antiferromagnetic magnons have emerged. Antiferromagnetic magnons can exhibit the full range of circular to linear polarization in collinear antiferromagnets[5], and a finite magnon Hall angle is predicted in chiral antiferromagnets[6]. Theoretical work has also predicted the interaction between antiferromagnetic magnons and spin-textures[7,8], by respective changes of their polarization and of the local Néel order. Ballistic[9], diffusive[10] and spin-superfluid regimes through magnon condensation[11,12] have been predicted, and electrical signatures by spin-orbit coupling effects are expected[13–15].

Experimental observations of this rich physics have started to emerge, with recent reports of long-distance spin-transport near room temperature in the easy-axis phase of hematite[8,16] and at low temperatures in antiferromagnetic quantum Hall graphene[17]. However, the complex spin transport features in collinear antiferromagnets are generally not indicative of the coherent transport regime although signatures have recently been claimed[18]. Furthermore, while the transport in easy-axis AFMs that is expected from the circular polarization of the magnons has been clearly observed[16], the possibility to propagate long-distance spin-currents in the wide-spread collinear easy-plane antiferromagnets remains an open question in the emerging field of antiferromagnetic magnonics[10]. Lastly, achieving long-distance room temperature spin-transport has not been achieved yet which is a prerequisite to integrate antiferromagnets in spintronic and magnonic devices.

---

* corresponding authors: romain.lebrun@cnrs-thales.fr ; klaeui@uni-mainz.de

Hematite, $\alpha$-Fe$_2$O$_3$, is a model system to investigate the spin transport regime of easy-axis antiferromagnets as we recently reported[8,16] but the easy-axis phase is only present at low temperatures. Above the Morin temperature ($T_{Morin}$ = 260 K), undoped hematite single crystals undergo a transition from an easy-axis to an easy-plane AFM, due to a change of sign of its anisotropy field $H_A$[19], with a small sub-lattice canting due to its internal Dzyaloshinskii-Moriya field[20]. One must notice that the Morin transition can disappear due to size effects in thin films and be recovered through doping[21]. A similar transition towards a canted easy-plane phase can be obtained at lower temperatures for sufficiently high fields in the spin-flop state[20,22]. In order to realize room temperature spin-transport, one needs to demonstrate the transport in the easy-plane phase. Hematite therefore represents a model system to simultaneously address and compare the origins of the magnonic transport in easy-axis and canted easy-plane antiferromagnets by making use of temperature and field cycling.

In this paper, we demonstrate that the easy-plane phase of the antiferromagnet hematite can transport spin information over long distances at room temperature. As a function of temperature, the spin transport length scale drops continuously. When going across the Morin transition there is no abrupt change but rather the transport length scale continuously changes with temperature. We associate this surprising behavior with the current induced correlated magnon pairs with a small difference of **k** vectors in combination with the ultra-low magnetic damping of hematite that we measure using electron paramagnetic resonance at frequencies of hundreds of gigahertz. Together we can explain the long-distance transport present in both the easy-axis and easy-plane phases and at elevated temperatures as required for applications.

## Results

### Spin transport through the Morin transition

To study the role of the antiferromagnetic symmetry and anisotropy in the transport of antiferromagnetic magnons, we performed nonlocal measurements on a crystal of the antiferromagnet hematite using platinum stripes, parallel to the projection of the in-plane projection of the easy-axis (along the **x** axis, sketch in **Fig. 1.a**). To measure magnon transport, we inject a charge current through the Pt injector, which generates a transverse spin current due to the spin Hall Effect (SHE). An electron spin accumulation builds up at the Pt/$\alpha$-Fe$_2$O$_3$ interface (along **y**) resulting in the excitation of spin-polarized magnons for a parallel alignment of the antiferromagnetic order and the interfacial electron spin accumulation. This non-equilibrium magnon population then diffuses away from the injector and is then absorbed by an electrically isolated Pt detector some distance away (0.5 to 10 μm). It is then converted to a charge current via the inverse SHE. This spin-bias signal can then be expressed as a nonlocal voltage $V_{el}$ as previously established[16].

While transport has so far been confined to the low temperature easy-axis phase, here we investigate the temperature dependence of the spin-transport signal through the Morin transition as shown in **Fig. 1.b**. As we previously reported[16], we observe at all temperatures below the Morin temperature ($T_M$ = 260 K) a peak of the spin-transport signal at the spin-flop field when the applied field leads to the Néel vector reorientation (**n // y**) and the softening of the magnetic systems closes the magnon gap. This divergence is less pronounced at lower temperatures for which the magnetic susceptibility and the spin-flop field are larger (about 8 T at 150 K[22]). The absence of detectable signal below 75 K indicates a diffusive transport process dominated by thermal magnons and no dominating spin superfluidity. At temperatures above $T_M$, in the easy-plane phase, the peak is less pronounced and the amplitude of the signal decreases at seen in **Fig. 1.c**.

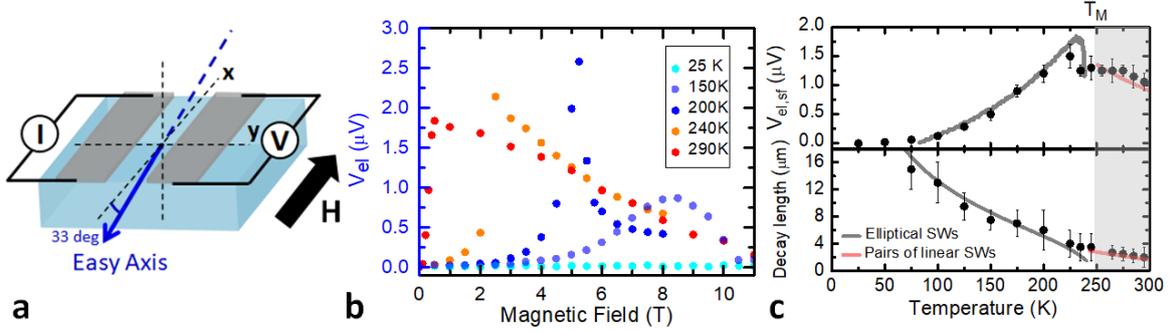

**Figure 1 Spin-transport through the Morin transition ($T_M$)** (a) Schematic of the nonlocal geometry of two electrically isolated Pt wires parallel to the in-plane projection of the easy axis. (b) Temperature dependence of the nonlocal spin-signals for a magnetic field parallel to the platinum stripes (33 deg from the easy-axis). Error bars from counting statistic are smaller than the symbol size. (c) Top: The spin signal $V_{el}$ is measured at the spin-flop field and approaches zero at low temperature, indicating a diffusive regime. Data obtained for an inter-stripe distance of 500 nm. Bottom: Spin-wave decay length (spin diffusion length λ for T < $T_M$ and dephasing length *L* for T > $T_M$) as a function of temperature for *H* applied along x at the spin-flop field. (Gray and red lines respectively correspond to fits with a magnon transport based on elliptically polarized spin-waves (SWs) and on pairs of linearly polarized spin-waves. The grey line would go to zero above the Morin transition due to the absence of elliptical spin-waves. For modelling we used the following data: exchange field $H_{ex}$ = 1040 T, $H_{DMI}$ =2.72 T, $H_{an,in}$ =24 µT as in Ref. [23]).

In parallel, we also measure a reduction of the magnon spin-diffusion length λ as the temperature increases. This decrease is in contrast to the increase with temperature observed in ferrimagnet YIG[24]. We detect a spin-transport signal for distances larger than 500 nm between the injector and the detector up to 320 K allowing us to determine that the spin-transport length scales even above room temperature are still in the range of µm. These features highlight the change of the spin-transport properties of diffusive magnons between the easy-plane and easy-axis antiferromagnetic phases.

## Spin transport in the canted easy-plane phase

To characterize the detailed magnon transport properties above the Morin temperature, and in particular identify whether the spin-current is carrier by the Néel vector or the weak canted moment that is orthogonal, we present in **Fig. 2** the angular and field dependences of the spin-signal in the canted easy-plane phase (sketch of **Fig. 2.a**). When we apply a field along the **x** or the **z** axis, the Néel vector smoothly reorients within the easy-plane and orients perpendicular to the stripes, i.e along **y**, and saturates at a field of about 0.4 T. This small spin-flop field in the easy-plane arises from magneto-elastic interactions, which emerge above the Morin temperature[25]. This spin-flop field, associated with a 6th order anisotropy term, leads to a threefold symmetry in the easy-plane and to a non-zero frequency gap[22]. Such threefold symmetry prevents a full compensation of the anisotropy fields, and is detrimental to achieve potential spin-superfluid regimes in linear response[12].

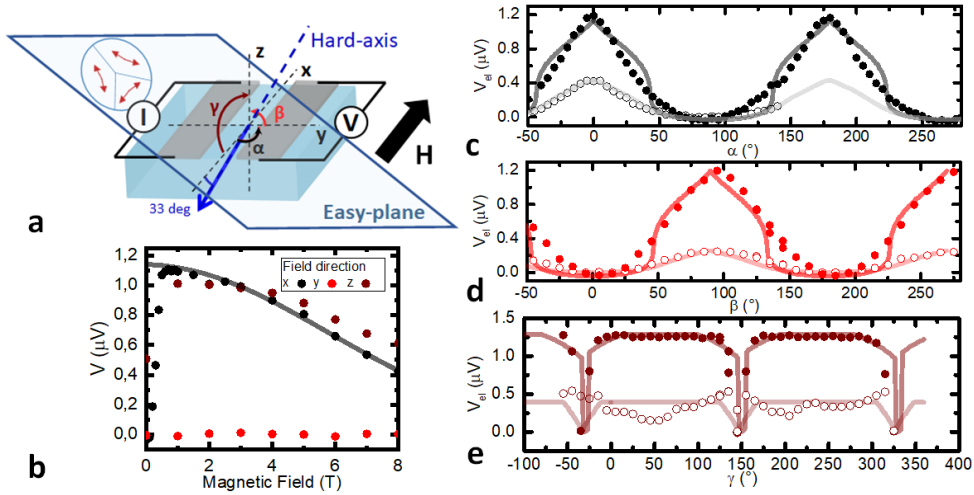

**Figure 2 Spin-transport in the easy-plane phase** (a) Sketch of the x-oriented devices relatively to the (111) easy-plane. α is the (xy) angle between the applied field and the x-axis. β is the (yz) angle between the applied field and the y-axis. γ is the (xz) angle between the applied field and the x-axis. (b) Spin transport signal for fields along the **x**, **y** and **z** directions (black line corresponds to a fit based on phenomenological model, described in the text, based on correlated pairs of linear spin-waves). (c, d, e) Spin transport signal in the α, β and γ planes. Filled and open symbols correspond to an applied field of 0.5 and 8 T respectively (shaded lines correspond to fits based on the phenomenological model based on pairs of linear spin-waves. For modelling we used the following data: exchange field $H_{ex}$ = 1040 T, $H_{DMI}$ =2.72 T, $H_{an,in}$ =24 µT as in Ref. [23]). Error bars from counting statistic are smaller than the symbol size.

    Above the spin-reorientation transition, the amplitude of the spin-signal smoothly decreases when the field is applied along the **x** and **z** axes. If instead, the field is applied perpendicular to the stripes, the Néel order orients perpendicular to the electron spin accumulation and no signal is observed (red curve in **Fig. 2.b**). This indicates that the spin information is transported along the Néel order direction and not by the weak canted moment; the transport is thus of antiferromagnetic nature as found previously for the easy-axis phase. The moment due to the canting of the sub-lattices plays no significant role here. We confirm these observations using the angular dependence in the (xy), (yz) and (xz) planes as shown in **Figs 2.c-2.e**. The transport signal shows a maximum for **H** parallel to either **x** or **z**, whilst a minimum is observed for a magnetic field applied along **y**. At 0.5 T, the signal is nearly always maximal in the γ-plane (**Fig. 2.e**) except at γ = -35 ± 5 ° (mod. 180 °) for which the field is applied perfectly along the hard-axis (c-axis). In this latter case, the condition **n** parallel to the current polarization **y** is not fulfilled and no spin current propagates. Considering the angular dependence of the signals shown in **Fig. 2.c-e**, one can see that the easy-plane symmetry plays a crucial role in the properties of the spin-transport signal[26]. In the (xy) and (yz) planes, the oscillations keep their shape at 8 T but their amplitude strongly decreases as expected from the measurements using a single field direction shown in **Fig. 2.b**. The increase of the externally applied magnetic field has two main effects: first, the increasing field enhances the magnon gap, indicating that low energy magnons with small **k** vectors dominate the spin-transport signal. Secondly, it modifies the magnon polarization; above the Morin transition, the ellipticity of the magnons near the center of the Brillouin zone[5], evolves towards a linear polarization with increasing temperature (due to the continuous increase in the hard axis anisotropy).

## Antiferromagnetic resonance and magnetic damping of hematite

Having established the possibility of spin transport in both the easy-axis and easy-plane phase of hematite, we need to understand the origin of the record spin-transport distances found in hematite. A key parameter of the magnon decay length in both easy-axis and easy-plane antiferromagnet is the magnetic damping. To obtain information about the magnetization dynamics of hematite and on this key parameter, we investigate the magnetization dynamics on a single crystal of hematite using magnetic resonance measurements from 120 GHz to 380 GHz[27]. From the dynamics of the probed uniform mode, we can extract information about the low **k** magnons which dominate the spin transport. In **Fig. 3.a-b**, we show the frequency and linewidth dependence of the low frequency mode. The frequency dependence of the mode can be fitted using our calculations[23]. Due to a wavelength smaller than the thickness of the crystal at high frequencies, we observe a broad peak at low magnetic field and multiple resonance peaks at higher magnetic fields (see inset of **Fig. 3.a**). At magnetic fields below 4 T, the presence of magnetostatic modes leads to additional linewidth broadening that prevents us from extracting the magnetic damping (see Methods). Above this value, we can extract the resonance linewidth by measuring the average peak-to-peak distance of the resonances. This technique leads to large error bars as shown in **Fig. 3.b**, but the results are in agreement with previous measurements using neutron[2], terahertz[28] and electron paramagnetic resonance[29,30] spectroscopy. The results cannot be fitted well with the existing simple theories of antiferromagnetic resonances[31], but we can deduce a magnetic damping with an upper limit of $10^{-5}$. This indicates that the magnetic damping of hematite α is of the same order as for YIG, the ferromagnetic material with so far the lowest reported magnetic damping of any magnetic compound.

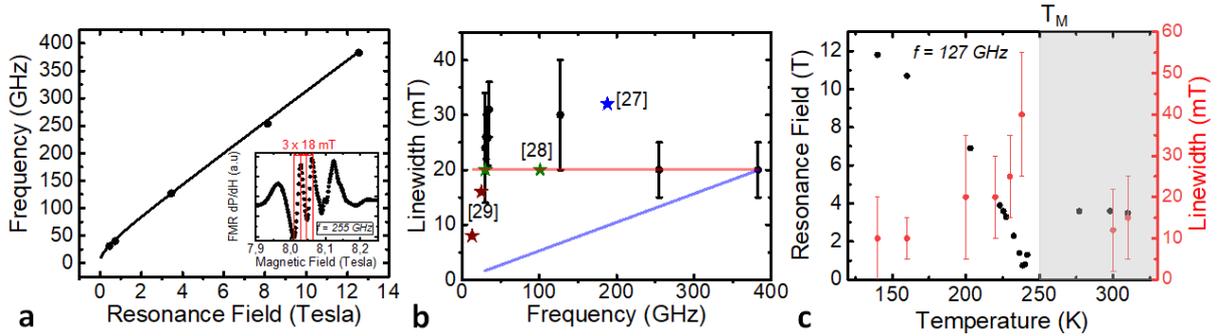

**Figure 3 Magnetic resonance and relaxation** (a) Resonance frequency as a function of magnetic field for the low frequency mode of hematite at room temperature. Inset: Resonance peak at 255 GHz for a 0.5 mm thick single crystal. (b) Linewidth as a function of frequency. The blue, brown and green points correspond to literature values from Refs. [28–30]. Shadow red and blue lines correspond to theoretical calculations from two models from Fink et al.[31] with (α = $10^{-6}$) and without (α = $10^{-5}$) a dependence of AFM linewidth on the anisotropy. (c) Resonance field and linewidth as a function of temperature for an excitation frequency of 127 GHz. Error bars correspond to peak statistics.

We also performed magnetic resonance measurements for a fixed excitation frequency of 127 GHz as a function of temperature as shown in **Fig. 3.c**. First, we observe that the linewidths at 200 K and 300 K are of similar order of magnitude showing that the magnetic damping is low in both the easy-plane and easy-axis phases. We also observe a small increase of the linewidth around $T_M$, indicative of stronger dissipation processes at the transition which could arise from the minimum anisotropy at the Morin transition[31].

## Discussion

To understand the observed spin transport resulting from the magnon properties in hematite, we develop a simple phenomenological model of magnon transport which defines a phase diagram with two regions (see **Fig. S1** in Supplementary Information[32]). Below the critical field $H_{cr}(T)$ for the spin flop, the equilibrium orientation of the Néel vector $\mathbf{n}^{(0)}(\mathbf{H})$ varies depending on the magnetic field $\mathbf{H}$ (inset in **Fig. S1** in Supplementary Information[32]). In this region (both above and below the Morin point) the magnon modes are polarized parallel or antiparallel to equilibrium orientation of the Néel vector. Spin transport in this region is similar to spin transport of uniaxial antiferromagnets discussed in Refs.[9,16]. Above the critical field $H_{cr}(T)$, the Néel vector $\mathbf{n}^{(0)}(\mathbf{H}) \parallel \hat{y} \perp \mathbf{H}$ is oriented perpendicular to the magnetic field, the magnon eigenmodes in the absence of a spin-current are linearly polarized. So, to understand the observed spin-transport signal, we need to discuss the magnon spin transport in antiferromagnets with linearly polarized magnon modes.

To address this, we first analyze the magnon spectrum in the presence of spin polarized currents emerging from the current distribution in the Pt injector electrode. Below the critical field $H_{cr}(T)$ where the eigenmodes have an elliptical polarization, i.e carry spin information, the current-induced anti-damping torque suppresses (enhances) the damping of the magnons polarized parallel (antiparallel) to the spin of current[33]. According to the fluctuation dissipation theorem[34], this can be interpreted as a splitting of the effective temperature $T_\pm$ for spin-up and spin-down magnons[35] and lead to the creation of a nonequilibrium spin-accumulation of magnons (see Supplementary Information[32]):

$$\boldsymbol{\mu} \propto \boldsymbol{n}^{(0)}(\mathbf{H}_{\text{curr}} \cdot \mathbf{n}^{(0)}) \left[ s_+ f\left(\frac{\omega_+}{T}\right) + s_- f\left(\frac{\omega_-}{T}\right) \right], \qquad (1)$$

where $\mathbf{H}_{\text{curr}}$ is the effective field parallel to the electron spin accumulation and proportional to the current density $j$, $f$ is the equilibrium Bose-Einstein distribution function and $\omega_\pm$ is the magnon frequency. The spin polarization of the magnon mode $0 \leq s_\pm \leq 1$ is related to its ellipticity[5,33], and depends on the magnetic field: $s_\pm \propto \mathbf{H} \cdot \mathbf{n}^{(0)}$. The non-monotonic field dependence of the voltage $V(H) \propto \mu_y$ shown in **Fig. 1.b.** is thus explained by field-induced variation of the ellipticity $s_\pm$ and the rotation of the Néel vector. This model also predicts the growth of the $V(H)$ maximum with temperature once the temperature dependence of the magnetic easy-axis anisotropy $H_{an}(T)$[23] is taken into account, as shown in the top panel of **Fig. 1.c.** by the grey line.

Above the critical field $H_{cr}(T)$, below and above the Morin transition, the situation changes completely: the magnon modes are linearly polarized and the current-induced torques establish correlations between the linearly-polarized magnon modes with orthogonal polarizations[36]. The pairs of two linearly polarized magnons with different frequencies, though coupled by current, carry no spin angular momentum and do not contribute to spin transport. However, the pairs, whose wave vectors $\mathbf{k}$ satisfy the energy conservation relation, $\omega_1^2 + c^2 \mathbf{k}_1^2 = \omega_2^2 + c^2 \mathbf{k}_2^2$, (see **Fig. 4**), generate a net nonequilibrium magnon spin accumulation (1) with $\omega_+ \to \sqrt{\omega_1^2 + c^2 \mathbf{k}_1^2}$, $s_+ = 1$, $s_- = 0$, where $H_{curr}(\mathbf{k}_1 - \mathbf{k}_2)$ corresponds to the space Fourier component of the current j($\mathbf{k_1}$ - $\mathbf{k_2}$), $\omega_{1,2}$ to the gaps ($\mathbf{k}$ = 0) of the high frequency and low frequency magnon branches, c is the limiting velocity of magnons. These pairs thus carry spin-information, which explains the presence of a non-zero spin transport signal above the spin-flop field and in the easy-plane phase above the Morin transition.

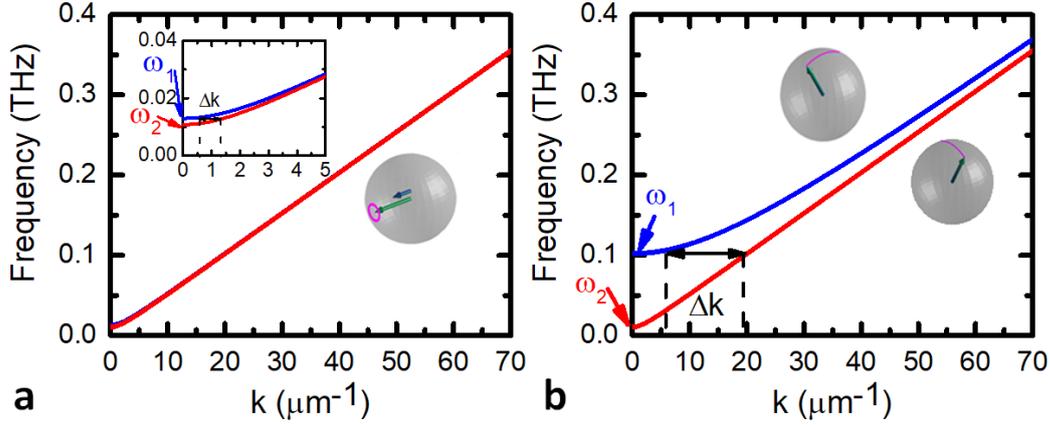

**Figure 4 Magnon dispersion curves of the two magnon modes of an easy-plane antiferromagnet** (a) low (close to the Morin transition) and (b) high anisotropy (far from the Morin transition, here + 30 K [29]). The pairs of magnons which satisfy energy conservation have small (large) Δk for low (high) anisotropy. For the calculations, we use a limiting magnon velocity c = 5.10⁴ m/s as estimated from the magnon dispersion of Martin et al.[37]

Below the Morin transition, the spin propagation length $\lambda$ of magnons was measured at the spin-flop field, i.e at the phase boundary $H_{cr}(T)$ where the signal $V_{el}$ reaches a maximal value (**Fig. 1.b**). The corresponding temperature dependence can be well fitted with the law $\lambda \propto \sqrt{H_{an}(T)/T}$ which correlates with the magnon spin-diffusion length $\lambda_T \propto 1/\sqrt{T}$ [9] as shown in the bottom panel of **Fig. 1.c**. The additional factor $\sqrt{H_{an}(T)}$ can be attributed to the effect of the magnetic field, which stabilizes elliptically polarized states and whose value at the phase boundary $H_{cr}(T)$ scales with $H_{an}$.

In the region above $H_{cr}(T)$ and above the Morin transition, the characteristic decay length of the magnon is dominated by the dephasing-induced attenuation of the signal. This is also illustrated in **Fig. 2.b**, which shows the field dependence $V(H)$ at a fixed distance $x$ from the injector electrode. Formally, the spatial dependence of the attenuated signal in the presence of dephasing follows the same exponential decay $V(y) \propto \exp\left(-\frac{y}{L}\right)$ as in the case of diffusion. However, the characteristic length $L$ depends on the difference $|\mathbf{k}_1 - \mathbf{k}_2|$ and, correspondingly, on the difference of magnon frequencies. We thus establish that the expression of the characteristic dephasing length is:

$$L = \frac{2\pi c^2 k_y}{\omega_1^2 - \omega_2^2}, (2)$$

where we assumed that $\Delta k_z \ll k_y$. One can notice that the magnetic field increases the splitting between the frequencies $|\omega_1^2 - \omega_2^2|$, which explain that $V(H)$ diminishes together with $L(H)$. Furthermore, close to the Morin transition temperature, the values of the in-plane ($H_{an\perp}$) and out-of-plane magnetic anisotropies ($H_{an}$) are of the same order of magnitude. As a consequence, $|\omega_1^2 - \omega_2^2| \propto |H_{an} - H_{an\perp}|$ is relatively small and $L$ is relatively large (in the μm range). It should however be noted that $H_{an\perp}$ strongly increases above the Morin transition whilst $H_{an}$ remains nearly constant[29]. Far above the Morin transition, the frequency splitting is so strong even in absence of the magnetic field, that the dephasing length is below the experimental resolution. This result highlights the importance of having inplane and out-of-plane anisotropies of the same order to propagate spin-information, which makes cubic antiferromagnets potential candidates in this purpose if they exhibit magnetic damping as low as hematite.

This long-distance spin-transport in both easy-axis and, in particular, easy-plane antiferromagnets and the observed ultra-low magnetic damping are remarkable features. Our findings broaden the class of materials in which one can use to propagate spin information. Not only easy-axis antiferromagnets with intrinsic circularly polarized magnon modes can carry spin-information but also in easy-plane antiferromagnets one can electrically generate pairs of linearly polarized spin-waves, which carry an effective circular polarization and thus a spin-information. The dephasing length of these magnon pairs is strongly dependent on the difference of their **k** vectors and thus on the magnetic anisotropies of the antiferromagnet. One can also control the **Δk** of the two magnon branches by applying a magnetic field or by varying the temperature. Secondly, the combined transport and antiferromagnetic resonance measurements highlight the high potential of low damping antiferromagnetic insulators, both with easy-axis and easy-plane anisotropies, for their integration into magnonic and spintronic devices. Our findings potentially open the transport also to hematite thin films, which are intrinsically mostly in the easy plane phase[38]. Significant further advances in the fabrication of high-quality thin films with large domains are necessary to realize future devices[8]. More generally, insulating antiferromagnets can have magnetic damping as low as the best ferromagnets and can also transport spin-information at room temperature over large length scales, which are both key features for magnonic devices.

**Acknowledgement**

R.L acknowledges the European Union's Horizon 2020 research and innovation program under the Marie Skłodowska-Curie grant agreements FAST number 752195. R.L., A.R. and M.K. acknowledge support from the Graduate School of Excellence Materials Science in Mainz (MAINZ) DFG 266, the DAAD (Spintronics network, Project No. 57334897) and all groups from Mainz acknowledge that this work was funded by the Deutsche Forschungsgemeinschaft (DFG, German Research Foundation) -SFB TRR 173 –268565370 (projects A01, A03, A11, B02, B11 and B12). R.L. and M.K. acknowledge financial support from the Horizon 2020 Framework Programme of the European Commission under FET-Open grant agreement no. 863155 (s-Nebula). R.L., A.R., and M.K. acknowledge support from the DFG project number 423441604. O.G. and J.S. acknowledge the Alexander von Humboldt Foundation, the ERC Synergy Grant SC2 (No. 610115). This work was also supported by the Max Planck Graduate Centre (MPGC). A.Q, A.B., M.K. were supported by the Research Council of Norway through its Centres of Excellence funding scheme, project number 262633 "QuSpin". V.B acknowledges financial support from the French national research agency (ANR) (Grant Number ANR-15-CE24-0015-01) and the bottom-up exploratory program of the CEA (Grant Number PE-18P31-ELSA).

**Competing interests**

The authors declare no competing interests.

**Author contributions**

R.L. and M.K proposed and supervised the project. R.L and A.R. performed the transport experiments. R.L, U.E, V.B and A.-L.B performed the magnetic resonance measurements. A.R patterned the samples. R.L, O.G, A.R. analysed the data. O.G. performed the analytical calculations with inputs from R. L., M. K., A.Q., J. S. and A.B.. R.L, O. G, A.R. and M.K wrote the paper. All authors commented on the manuscript.

**Data availability**

The data that support the findings of this study are available from the corresponding authors upon reasonable request. Correspondence and requests for materials should be addressed to R.L. or M.K.

**References**

1. Hutchings, M. T. & Samuelsen, E. J. Measurement of Spin-Wave Dispersion in NiO by Inelastic Neutron Scattering and Its Relation to Magnetic Properties. *Phys. Rev. B* **6**, 3447–3461 (1972).
2. Samuelsen, E. J. & Shirane, G. Inelastic neutron scattering investigation of spin waves and magnetic interactions in α-Fe2O3. *Phys. Status Solidi B* **42**, 241–256 (1970).
3. Baltz, V. *et al.* Antiferromagnetic spintronics. *Rev. Mod. Phys.* **90**, 015005 (2018).
4. Jungwirth, T., Marti, X., Wadley, P. & Wunderlich, J. Antiferromagnetic spintronics. *Nat. Nanotechnol.* **11**, 231–241 (2016).
5. Rezende, S. M., Azevedo, A. & Rodríguez-Suárez, R. L. Introduction to antiferromagnetic magnons. *J. Appl. Phys.* **126**, 151101 (2019).
6. Mook, A., Göbel, B., Henk, J. & Mertig, I. Magnon transport in noncollinear spin textures: Anisotropies and topological magnon Hall effects. *Phys. Rev. B* **95**, 020401 (2017).
7. Tveten, E. G., Qaiumzadeh, A. & Brataas, A. Antiferromagnetic Domain Wall Motion Induced by Spin Waves. *Phys. Rev. Lett.* **112**, 147204 (2014).
8. Ross, A. *et al.* Propagation Length of Antiferromagnetic Magnons Governed by Domain Configurations. *Nano Lett.* **20**, 306–313 (2020).
9. Bender, S. A., Skarsvåg, H., Brataas, A. & Duine, R. A. Enhanced Spin Conductance of a Thin-Film Insulating Antiferromagnet. *Phys. Rev. Lett.* **119**, 056804 (2017).
10. Rezende, S. M., Rodríguez-Suárez, R. L. & Azevedo, A. Diffusive magnonic spin transport in antiferromagnetic insulators. *Phys. Rev. B* **93**, 054412 (2016).
11. Takei, S., Halperin, B. I., Yacoby, A. & Tserkovnyak, Y. Superfluid spin transport through antiferromagnetic insulators. *Phys. Rev. B* **90**, 094408 (2014).
12. Qaiumzadeh, A., Skarsvåg, H., Holmqvist, C. & Brataas, A. Spin Superfluidity in Biaxial Antiferromagnetic Insulators. *Phys. Rev. Lett.* **118**, 137201 (2017).
13. Cheng, R., Xiao, J., Niu, Q. & Brataas, A. Spin Pumping and Spin-Transfer Torques in Antiferromagnets. *Phys. Rev. Lett.* **113**, 057601 (2014).
14. Johansen, Ø. & Brataas, A. Spin pumping and inverse spin Hall voltages from dynamical antiferromagnets. *Phys. Rev. B* **95**, 220408 (2017).
15. Li, J. *et al.* Spin current from sub-terahertz-generated antiferromagnetic magnons. *Nature* **578**, 70–74 (2020).
16. Lebrun, R. *et al.* Tunable long-distance spin transport in a crystalline antiferromagnetic iron oxide. *Nature* **561**, 222 (2018).
17. Stepanov, P. *et al.* Long-distance spin transport through a graphene quantum Hall antiferromagnet. *Nat. Phys.* 1 (2018) doi:10.1038/s41567-018-0161-5.
18. Yuan, W. *et al.* Experimental signatures of spin superfluid ground state in canted antiferromagnet Cr2O3 via nonlocal spin transport. *Sci. Adv.* **4**, eaat1098 (2018).
19. Artman, J. O., Murphy, J. C. & Foner, S. Magnetic Anisotropy in Antiferromagnetic Corundum-Type Sesquioxides. *Phys. Rev.* **138**, A912–A917 (1965).
20. Moriya, T. Anisotropic Superexchange Interaction and Weak Ferromagnetism. *Phys. Rev.* **120**, 91–98 (1960).
21. Ellis, D. S. *et al.* Magnetic states at the surface of alpha-Fe2O3 thin films doped with Ti, Zn, or Sn. *Phys. Rev. B* **96**, 094426 (2017).
22. Elliston, P. R. & Troup, G. J. Some antiferromagnetic resonance measurements in α-Fe2O3. *J. Phys. C Solid State Phys.* **1**, 169 (1968).
23. Lebrun, R. *et al.* Anisotropies and magnetic phase transitions in insulating antiferromagnets determined by a Spin-Hall magnetoresistance probe. *Commun. Phys.* **2**, 50 (2019).
24. Cornelissen, L. J., Shan, J. & van Wees, B. J. Temperature dependence of the magnon spin diffusion length and magnon spin conductivity in the magnetic insulator yttrium iron garnet. *Phys. Rev. B* **94**, 180402 (2016).
25. Liebermann, R. C. & Banerjee, S. K. Magnetoelastic interactions in hematite: Implications for geophysics. *J. Geophys. Res.* **76**, 2735–2756 (1971).
26. Wang, H. *et al.* Antiferromagnetic anisotropy determination by spin Hall magnetoresistance. *J. Appl. Phys.* **122**, 083907 (2017).

27. Barra, A. L., Gatteschi, D. & Sessoli, R. High-frequency EPR spectra of a molecular nanomagnet: Understanding quantum tunneling of the magnetization. *Phys. Rev. B* **56**, 8192–8198 (1997).
28. Chou, S. G. *et al.* High-Resolution Terahertz Optical Absorption Study of the Antiferromagnetic Resonance Transition in Hematite (α-Fe2O3). *J. Phys. Chem. C* **116**, 16161–16166 (2012).
29. Velikov, L. V. & Rudashevskii, E. G. Antiferromagnetic resonance in hematite in the weakly ferromagnetic state. *Zh Eksp Teor Fiz* **56**, 1557–1564.
30. Searle, C. W. & Wang, S. T. Magnetic-Resonance Properties of Pure and Titanium-Doped Hematite. *J. Appl. Phys.* **39**, 1025–1026 (1968).
31. Fink, H. J. Resonance Line Shapes of Weak Ferromagnets of the α-Fe2O3 and NiF2 Type. *Phys. Rev.* **133**, 1322–1326 (1964).
32. See Supplementary Information for the details on the analytical model describing the transport of the spin-wave transport and spin-transport data for other inter-stripe distances and temperatures.
33. Gomonay, O., Yamamoto, K. & Sinova, J. Spin caloric effects in antiferromagnets assisted by an external spin current. *J. Phys. Appl. Phys.* **51**, 264004 (2018).
34. Kovalev, A. A. & Tserkovnyak, Y. Thermomagnonic spin transfer and Peltier effects in insulating magnets. *EPL Europhys. Lett.* **97**, 67002 (2012).
35. Gomonay, O & Loktev, V. Spin torque antiferromagnetic nanooscillator in the presence of magnetic noise. *Condens. Matter Phys.* **15**, 43703 (2012).
36. Tatara, G. & Pauyac, C. O. Theory of spin transport through an antiferromagnetic insulator. *Phys. Rev. B* **99**, 180405 (2019).
37. Martin, T. P., Merlin, R., Huffman, D. R. & Cardona, M. Resonant two magnon Raman scattering in α-Fe2O3. *Solid State Commun.* **22**, 565–567 (1977).
38. Fischer, J. *et al.* Large Spin Hall Magnetoresistance in Antiferromagnetic α − Fe 2 O 3 / Pt Heterostructures. *Phys. Rev. Appl.* **13**, 014019 (2020).
39. Duncan, J. A., Storey, B. E., Tooke, A. O. & Cracknell, A. P. Magnetostatic modes observed in thin single-crystal films of yttrium iron garnet at Q-band frequencies. *J. Phys. C Solid State Phys.* **13**, 2079–2095 (1980).
40. Beeman, D. E. Magnetostatic Modes in Antiferromagnets and Canted Antiferromagnets. *J. Appl. Phys.* **37**, 1136–1137 (1966).